\documentclass[preprint,showpacs,preprintnumbers,amsmath,amssymb]{revtex4}

\usepackage{graphicx}
\usepackage{epsfig}
\usepackage{bm}
\usepackage{amsfonts}
\usepackage{subfigure}

\begin{document}

\title[Brans-Dicke inflation]{Brans-Dicke inflation in light of the Planck 2015 data}

\author{B. Tahmasebzadeh$^{1}$\footnote{b.tahmasebzadeh@iasbs.ac.ir}, K. Rezazadeh$^{2}$\footnote{rezazadeh86@gmail.com} and K. Karami$^{2}$\footnote{kkarami@uok.ac.ir}}

\address{$^1$Department of Physics, Institute for Advanced Studies in Basic Sciences (IASBS), P.O. Box 45195-1159, Zanjan, Iran\\
$^2$Department of Physics, University of Kurdistan, Pasdaran
St., P.O. Box 66177-15175, Sanandaj, Iran}

\begin{abstract}
We study inflation in the Brans-Dicke gravity as a special model of the scalar-tensor gravity. We obtain the inflationary observables containing the scalar spectral index, the tensor-to-scalar ratio, the running of the scalar spectral index and the equilateral non-Gaussianity parameter in terms of the general form of the potential in the Jordan frame. Then, we compare the results for various inflationary potentials in light of the Planck 2015 data. Our study shows that in the Brans-Dicke gravity, the power-law, inverse power-law and exponential potentials are ruled out by the Planck 2015 data. But, the hilltop, Higgs, Coleman-Weinberg and natural potentials can be compatible with Planck 2015 TT,TE,EE+lowP data at 95\% CL. Moreover, the D-brane, SB SUSY and displaced quadratic potentials can be in well agreement with the observational data since their results can lie inside the 68\% CL region of Planck 2015 TT,TE,EE+lowP data.
\end{abstract}



\maketitle


\section{Introduction}\label{sec:int}

The model of Hot Big Bang cosmology has impressive successes such as explaining the light nucleosynthesis and the cosmic microwave background (CMB) radiation. Despite its considerable successes, it suffers from central problems such as the flatness problem, the horizon problem and also the magnetic monopole problem. Inflation theory was proposed to solve all of these problems \cite{Starobinsky1980, Sato1981a, Sato1981, Guth1981, Linde1982, Albrecht1982, Linde1983}. Inflation is not a replacement for the Hot Big Bang cosmology, but rather an extra add-on idea which supposes that a short period of rapid accelerated expansion has occurred before the radiation dominated era. In addition to solving the problems of the Hot Big Bang cosmology, inflation can provide a reasonable explanation for the anisotropy observed in the CMB radiation and also in the large-scale structure (LSS) of the universe \cite{Mukhanov1981, Hawking1982, Starobinsky1982, Guth1982}. This fact makes it possible for us to contact the late time observations to the early stages of our universe.  Important observational results are provided by the Planck satellite from probing of the CMB radiation anisotropies in both temperature and polarization \cite{Planck2015}. Using these observational results, we can distinguish viable inflationary models and also constrain them.

The standard inflationary scenario is based on a canonical scalar field in the framework of Einstein gravity. Viability of different inflationary models in the framework of standard inflationary scenario in light of the observational results has been extensively investigated in the literature \cite{Martin2014a, Martin2014b, Rezazadeh2015, Huang2015, Okada2014}. So far, many inflationary models have been proposed. One important class of inflationary models are based on the extended theories of gravity. The well-known instance for this class of models is the Starobinsky $R^2$ inflation \cite{Starobinsky1980}. Despite the fact that this model is the first inflationary model, it is in well agreement with the observational results \cite{Planck2015, Martin2014a, Martin2014b, Rezazadeh2015, Huang2015}. Inflationary models on the extended theories of gravity have been extensively studied in the literature \cite{Felice2011a, Tsujikawa2013, Artymowski2014, Artymowski2015, Barrow1995, Berman2009, Rinaldi2015, Sebastiani2015, Myrzakulov2015, Myrzakul2015, Nashed2014, Jamil2015, Sharif2015, Rezazadeh2016, Bamba2014, Bamba2014a, Bamba2014b, Bamba2015, Myrzakulov2015a, Sebastiani2014, Cerioni2009, Cerioni2010, Finelli2008, Tronconi2011, Kamenshchik2011, Kumar2016, Kannike2015, Kannike2016, Marzola2016}.

One important branch of the extended theories of gravity is the scalar-tensor gravity which is a general theory that includes the $f(R)$-gravity, the Brans-Dicke gravity and the dilatonic gravity \cite{Faraoni2004, Fujii2004, Felice2010}. In the present paper, we focus on the Brans-Dicke gravity and study inflation in this framework.

In study of inflation, the scalar field is called ``inflaton'' that can provide a negative pressure needed to have an accelerated expansion. During inflation, the inflaton rolls slowly downward a potential and we can examine its evolution classically \cite{Lyth2009, Baumann2009}. At the end of inflation, the inflaton begins to oscillate around the minimum of the potential that leads to particle production and provides for the universe to transit into the radiation dominated era. This period is known as the ``reheating'' process that its details are unknown to us so far. Also, we don't know the shape of the inflationary potential that determines the dynamics of the inflaton. In order to understand the inflationary potential, we need more advances in both theory and observations. However, by examination of different potentials in light of the observational results, we can specify some features of the original inflationary potential. In order to relate the present time observations to the inflationary era, we note that besides the classical evolution, the inflaton scalar field has some quantum fluctuations during inflation that can lead to the primordial perturbations whose we can see the imprints on the anisotropies observed in the CMB radiation and in the LSS formation \cite{Lyth2009, Baumann2009, Mukhanov1992, Mukhanov2005, Weinberg2008, Malik2009}.

Note that from the energy scale of the primordial universe, it is believed that cosmological inflation has occurred in the regime of high energy physics. Also in one hand, the effective quantum field theory predicts that the high energy theory has fields with non-canonical kinetic terms \cite{Franche2010, Franche2010a, Tolley2010}. On the other hand, from the action of the Brans-Dicke gravity, we know that the kinetic term of this theory has a non-canonical form. This motivates us to investigate the cosmic inflation of the early universe within the framework of the Brans-Dicke gravity. In our work, we concentrate on the various inflationary potentials which have motivations from quantum field theory or string theory and check their viability in light of the Planck 2015 observational results. To do so, first we present a brief review on the scalar-tensor gravity that it will be done in sec. \ref{sec:st}. Then, in sec. \ref{sec:BD}, we will apply the results of sec. \ref{sec:st} for the Brans-Dicke gravity as a special case of the scalar-tensor gravity and find the relations of the inflationary observables. This makes it possible for us to examine various inflationary potentials in comparison with the observational results that we proceed to it in sec. \ref{sec:pot}. Finally, in sec. \ref{sec:con}, we summarize our concluding remarks.

\section{A brief review on the scalar-tensor gravity}\label{sec:st}

At first, in 1950, Jordan applied a scalar field in the gravitational part of the action. Then, in 1961, Brans and Dicke \cite{Brans1961} introduced a formalism for gravity in which the metric field together with a scalar field have been invoked to describe the gravitational force. After the discovery of the present accelerated expansion of the universe in 1998, other models on the base of the scalar-tensor gravity were proposed to explain this phenomenon \cite{Amendola1999, Chiba1999, Uzan1999, Perrotta1999, Boisseau2000, Esposito2001, Torres2002}. In this class of models, a scalar field is considered to solve the cosmological constant problems. The scalar-tensor gravity relative to the other competitor theories, posses the advantage that it can involve the dark energy in the form of the energy-momentum tensor $T_{\mu \nu }^\varphi$ and involve the modified gravity in the form of the Einstein tensor $G_{\mu \nu }^\varphi$.

Throughout this paper we take the Jordan frame as the physical frame. In the Jordan frame, the general action of the scalar-tensor models can be written in the form \cite{Faraoni2004, Fujii2004, Felice2010a, Felice2010, Abdolmaleki2014}
\begin{equation}
S_{J} =  \int {{d^4}} x\sqrt { - g} \left[ {\frac{1}{2}f(R,\varphi ) - \frac{1}{2}\omega (\varphi ){g^{\mu \nu }}{\partial _\mu }\varphi {\partial _\nu }\varphi  - U(\varphi )} \right],\label{Sst}
\end{equation}
where $f(R,\varphi )$ is a general function of the Ricci scalar $R$ and the scalar field $\varphi$ while the parameter $\omega$ and the potential $U$ are general functions of $\varphi$. Hereafter, we take the reduced Planck mass equal to unity, ${M_P} \equiv 1/\sqrt {8\pi G}  = 1$. The above action for the scalar-tensor gravity includes the $f(R)$ models, the Brans-Dicke gravity and also the dilatonic models \cite{Faraoni2004, Fujii2004, Felice2010}.

Now, we turn to examine the dynamics of background cosmology in the scalar-tensor gravity. The variation of the action (\ref{Sst}) with respect to the metric tensor ${g_{\mu \nu }}$ leads to \cite{Felice2010, Abdolmaleki2014}
\begin{align}
\nonumber
& F{R_{\mu \nu }} - \frac{1}{2}f{g_{\mu \nu }} - {\nabla _\mu }{\nabla _\nu }F + {g_{\mu \nu }}{\nabla ^\alpha }{\nabla _\alpha }F =\\
\label{FRmunu}
& \omega (\varphi )\left( {{\nabla _\mu }\varphi {\nabla _\nu }\varphi  - \frac{1}{2}{g_{\mu \nu }}{\nabla ^\lambda }\varphi {\nabla _\lambda }\varphi } \right) - U(\varphi ){g_{\mu \nu }},
\end{align}
where ${\nabla _\mu }$ indicates covariant derivative and the function $F$ is defined as $F \equiv \partial f/\partial R$. Variation of the action (\ref{Sst}) relative to the scalar field $\varphi$ gives rise to
\begin{equation}\label{nnphi}
{\nabla ^\alpha }{\nabla _\alpha }\varphi  + \frac{1}{{2\omega (\varphi )}}\left( {{\omega _{,\varphi }}{\nabla ^\lambda }\varphi {\nabla _\lambda }\varphi  - 2{U_{,\varphi }} + {f_{,\varphi }}} \right) = 0,
‎\end{equation}
where ${U_{,\varphi }} \equiv dU/d\varphi$.

For a spatially flat Friedmann-Robertson-Walker (FRW) universe, Eqs. (\ref{FRmunu}) and (\ref{nnphi}) turn into \cite{Felice2010, Abdolmaleki2014}
\begin{eqnarray}
\label{3HF}
3{H^2}F - \frac{1}{2}(RF - f) + 3H\dot F - \frac{1}{2}\omega (\varphi ){\dot \varphi ^2} - U(\varphi ) &=& 0,
\\
\label{2FHdot}
\ddot F - H\dot F + 2F\dot H + \omega (\varphi ){\dot \varphi ^2} &=& 0,
\\
\label{varphiddot}
\ddot \varphi  + 3H\dot \varphi  + \frac{1}{{2\omega (\varphi )}}\left( {{\omega _{,\varphi }}{{\dot \varphi }^2} + 2{U_{,\varphi }} - {f_{,\varphi }}} \right) &=& 0,
\end{eqnarray}
where the dot denotes a derivative with respect to the cosmic time $t$. The Hubble parameter is denoted by $H \equiv \dot a/a$ where $a$ is the scale factor of the universe. Also, $R$ is the Ricci scalar which is given by
\begin{equation}\label{R}
R = 6\left( {\frac{{\ddot a}}{a} + \frac{{{{\dot a}^2}}}{{{a^2}}}} \right) = 6\left( {2{H^2} + \dot H} \right).
\end{equation}

In the following, we briefly review the cosmological perturbations in the scalar-tensor gravity (for more details about this subject see e.g. \cite{Hwang1990, Hwang1991, Hwang1997, Hwang2001a, Hwang2001, Hwang2002, Hwang2005, Felice2010}).
We consider a general perturbed metric about the flat FRW background as \cite{Felice2010}
\begin{equation}\label{FRWpert}
d{s^2} =  - (1 + 2\alpha )d{t^2} - 2a(t)({\partial _i}\beta  - {S_i})dt\,d{x^i} + {a^2}(t)({\delta _{ij}} + 2\psi {\delta _{ij}} + 2{\partial _i}{\partial _j}\gamma  + 2{\partial _j}{F_i} + {h_{ij}})d{x^i}d{x^j},
\end{equation}
where $\alpha$, $\beta$, $\psi$ and $\gamma$ are scalar perturbations, $S_i$ and $F_i$ are vector perturbations, and $h_{ij}$ is tensor perturbations. Also the energy-momentum tensor of a perfect fluid with perturbations is given by \cite{Felice2010}
\begin{equation}\label{Tmunu}
T_0^0 =  - (\rho  + \delta \rho ),\qquad T_i^0 =  - (\rho  + p){\partial _i}v,\qquad T_j^i = (p + \delta p)\delta _j^i,
\end{equation}
where $\rho$ and $p$ are the energy density and pressure of the perfect fluid, respectively. Also, ${\partial _i}v$ characterizes the scalar part of the velocity potential of the fluid. Here, it is useful to introduce the momentum density $\delta {q^i} \equiv ( \rho  +  p){v^i}$. Applying the scalar-vector-tensor (SVT) decomposition, the momentum density $\delta {q_i}$ can be expressed in terms of the scalar and vector parts, $\delta {q_i} = {\partial _i}\delta q + \delta {\hat q_i}$, where the vector part is divergenceless, ${\partial ^i}\delta {\hat q_i} = 0$. As a results, Eq. (\ref{Tmunu}) follows that the scalar part of the 3-momentum energy-momentum tensor $\delta T_i^0$ is equal to the scalar part of the momentum density ${\partial _i}\delta q$. The scalar part of the momentum density ${\partial_i}\delta q$ is used in definition of an important gauge-invariant quantity which is the curvature perturbation \cite{Felice2010}
\begin{equation}\label{calR}
{\cal R} \equiv \psi  + \frac{H}{{\rho  + p}}\delta q.
\end{equation}
Indeed, deviation from the homogeneous and isotropic FRW metric leads to perturbation in the constant-time spatial slices that this perturbation is specified by the curvature perturbation $\mathcal{R}$. The attractive feature of $\mathcal{R}$ is the fact that it remains constant outside the horizon. In particular, its amplitude is not affected by the unknown physical properties of the reheating process occurred at the end of inflation. It is the constancy of $\mathcal{R}$ outside the horizon that allows us to nevertheless predict cosmological observables. After inflation, the comoving horizon grows, so eventually all fluctuations will re-enter the horizon. After horizon re-entry, $\mathcal{R}$ determines the perturbations of the cosmic fluid resulting in the observed CMB anisotropies and the LSS \cite{Lyth2009, Baumann2009, Mukhanov1992, Mukhanov2005, Weinberg2008, Malik2009}.

In order to examine the evolution of $\mathcal{R}$ during inflation, first we need to write it in the form that remains invariant under coordinate transformation so that we can distinguish the physical perturbations from the nonphysical ones. Then, we should apply the Arnowitt-Deser-Misner (ADM) formalism based on the variation from the second order action to obtain the evolution equation for the curvature perturbation $\mathcal{R}$. Subsequently, we quantize the perturbations to find an initial condition for the evolution equation and then obtain its general solution for the quasi-de Sitter universe. In the next step, we evaluate the solution at the time of horizon exit and find the power spectrum of perturbations. In the standard inflationary model based on a minimally coupled scalar field in the Einstein gravity, the equation for the scalar perturbations is known as the ``Mukhanov-Sasaki equation''. For reviews on cosmological perturbations theory in the standard inflationary scenario see e.g. \cite{Lyth2009, Baumann2009, Mukhanov1992, Mukhanov2005, Weinberg2008, Malik2009}.

Using the perturbed equations in the scalar-tensor gravity, the equation of motion for the curvature perturbation can be derived as \cite{Felice2010}
\begin{equation}‎\label{d2u}
‎{u''_k} - \left( {{k^2} - \frac{{z''}}{z}} \right){u_k} = 0,
‎\end{equation}
where the prime indicates a derivative with respect to the conformal time $\tau  = \int {{a^{ - 1}}dt}$. The normalized variable $u$ is defined as
\begin{equation}‎\label{u}
‎u \equiv z \mathcal{R}.‎
‎\end{equation}
For the standard inflationary scenario, the variable $z$ is defined as $z \equiv a\dot \varphi /H$, but for the scalar-tensor gravity, this variable is obtained as
\begin{equation}‎\label{z}
‎z=a\sqrt{Q_{s}},
\end{equation}
where
\begin{equation}\label{Qs}
{Q_s} \equiv \frac{{\omega (\varphi ){{\dot \varphi }^2} + \frac{{3{{\dot F}^2}}}{{2F}}}}{{{{\left( {H + \frac{{\dot F}}{{2F}}} \right)}^2}}}.
\end{equation}
To obtain the power spectrum of the curvature perturbation, it is useful to introduce the slow-roll parameters \cite{Hwang2001a, Felice2010}
\begin{equation}\label{eps}
{\varepsilon _1} \equiv  - \frac{{\dot H}}{H^2}, ‎\hspace{.3cm} {\varepsilon _2} \equiv \frac{{\ddot \varphi }}{{H\dot \varphi }}‎, ‎\hspace{.3cm} {\varepsilon _3} \equiv \frac{{\dot F}}{{2HF}}, ‎\hspace{.3cm} {\varepsilon _4} \equiv \frac{{\dot E}}{{2HE}}‎‎.
\end{equation}
In the slow-roll approximation, we assume that the slow-roll parameters are much smaller than unity. In the above expressions, the parameter $E$ is defined as
\begin{equation}\label{E}
E \equiv F\left[ {\omega (\varphi ) + \frac{{3{{\dot F}^2}}}{{2{{\dot \varphi }^2}F}}} \right].
\end{equation}
Therefore, using Eqs. (\ref{Qs}), (\ref{eps}) and (\ref{E}), we can rewrite $Q_s$ as
\begin{equation}‎\label{Qs2}
‎Q_{s}=\dot{\varphi}^{2} \frac{E}{FH^{2}(1+\varepsilon_{3})^{2}}‎.
‎\end{equation}
If the slow-roll parameters are constant, i.e. ${\dot \varepsilon _i} = 0\,\, (i = 1,2,3,4)$, then using Eqs. (\ref{z}) and (\ref{Qs2}), we will have
\begin{equation}‎\label{d2z}
‎\frac{{z''}}{z} = \frac{{\nu _{\cal R}^2 - 1/4}}{{{\tau ^2}}},
‎\end{equation}
where
\begin{equation}‎\label{nuR}
\nu _{\cal R}^2 = \frac{1}{4} + \frac{{\left( {1 + {\varepsilon _1} + {\varepsilon _2} - {\varepsilon _3} + {\varepsilon _4}} \right)\left( {2 + {\varepsilon _2} - {\varepsilon _3} + {\varepsilon _4}} \right)}}{{{{\left( {1 - {\varepsilon _1}} \right)}^2}}}.‎
\end{equation}
In addition, the conformal time reads
\begin{equation}‎\label{tau}
‎\tau  =  - \frac{1}{{\left( {1 - {\varepsilon _1}} \right)aH}}.‎
\end{equation}
Consequently, the solution of Eq. (\ref{d2u}) can be expressed as a linear combination of the Hankel functions,
\begin{equation}‎\label{uktau}
‎{u_k}(\tau ) = \frac{{\sqrt {\pi |\tau |} }}{2}{e^{i(1 + 2{\nu _{\cal R}})\pi /4}}\left[ {{C_1}H_{{\nu _{\cal R}}}^{(1)}\left( {k|\tau |} \right) + {C_2}H_{{\nu _{\cal R}}}^{(2)}\left( {k|\tau |} \right)} \right],
‎\end{equation}
where the integration constants $C_1$ and $C_2$ are determined by imposing the suitable initial conditions. Finally, the acceptable solution for $u_{k}(\tau)$ is obtained as \cite{Felice2010}
\begin{equation}‎\label{uktau2}
‎{u_k}(\tau ) = \frac{{\sqrt {\pi |\tau |} }}{2}{e^{i(1 + 2{\nu _{\cal R}})\pi /4}}H_{{\nu _{\cal R}}}^{(1)}\left( {k|\tau |} \right)‎‎.
‎\end{equation}
The scalar power spectrum is defined as
\begin{equation}‎\label{Ps}
{{\cal P}_{s}} \equiv \frac{{{k^3}}}{{2{\pi ^2}}}|{\cal R}{|^2}‎.
‎\end{equation}
Using Eqs. (\ref{u}) and (\ref{uktau2}) in the above definition, we get
\begin{equation}‎\label{PsGam}
{{\cal P}_{s}} = \frac{1}{{{Q_s}}}{\left[ {\left( {1 - {\varepsilon _1}} \right)\frac{{\Gamma \left( {{\nu _{\cal R}}} \right)}}{{\Gamma \left( {3/2} \right)}}\frac{H}{{2\pi }}} \right]^2}{\left( {\frac{{|k\tau |}}{2}} \right)^{3 - 2{\nu _{\cal R}}}}‎,
‎\end{equation}
where $\Gamma$ is the Gamma function. The power spectrum of the curvature perturbation must be evaluated at the horizon crossing for which $k=aH$. In the slow-roll approximation, it takes the form
\begin{equation}\label{PsH}‎
‎{{\cal P}_{s}} \simeq \frac{1}{{{Q_s}}}{\left( {\frac{H}{{2\pi }}} \right)^2}\Big|_{k=aH}‎.
‎\end{equation}

The scale-dependence of the scalar power spectrum is specified by the scalar spectral index defined as
\begin{equation}‎\label{ns}
{n_s} - 1 \equiv \frac{{d\ln {{\cal P}_{s}}}}{{d\ln k}}.
‎\end{equation}
With the help of Eq. (\ref{PsGam}), the scalar spectral index (\ref{ns}) reads
\begin{equation}\label{nsnuR}‎
‎{n_s} - 1 = 3 - 2{\nu _{\cal R}}.
‎\end{equation}
In the slow-roll approximation, it therefore can be written as \cite{Felice2010}
\begin{equation}\label{nseps}
{n_s} \simeq 1 - 4{\varepsilon _1} - 2{\varepsilon _2} + 2{\varepsilon _3} - 2{\varepsilon _4}‎.
\end{equation}

Here, we concentrate on the tensor perturbations in the framework of the scalar-tensor gravity. The power spectrum of the tensor perturbations can be derived in a similar procedure to the one followed for the scalar perturbations and in the slow-roll regime it takes the form \cite{Felice2010}
\begin{equation}‎\label{Pt}
{{\cal P}_t} \simeq \frac{2}{{{\pi ^2}}}\frac{{{H^2}}}{F}\Big|_{k=aH}‎.
‎\end{equation}
To specify the scale-dependence of the tensor power spectrum, one can define the tensor spectral index
\begin{equation}\label{nt}
{n_t} \equiv \frac{{d\ln {{\cal P}_t}}}{{d\ln k}}.
\end{equation}
For the scalar-tensor gravity and in the slow-roll approximation, it can be obtained as
\begin{equation}‎\label{ntst}
{n_t} \simeq  - 2{\varepsilon _1} - 2{\varepsilon _3}.
\end{equation}
An important inflationary observable is the tensor-to-scalar ratio which is defined as
\begin{equation}‎\label{r}
‎r \equiv \frac{{\cal P}_t}{{\cal P}_{s}}‎.
‎\end{equation}
Using Eqs. (\ref{PsH}) and (\ref{Pt}) in (\ref{r}), the tensor-to-scalar ratio for the scalar-tensor gravity in the slow-roll approximation turns into
\begin{equation}\label{rst}
‎r \simeq 8\frac{{{Q_s}}}{F}.
\end{equation}

So far, we have obtained the inflationary observables in the Jordan frame which is our physical frame in this paper. Applying the conformal transformations, we can go from the Jordan frame to the Einstein frame and calculate the inflationary observables in that frame too. The issue of the conformal transformations is an important subject in the context of modified theories of gravity. Also, implications of the Einstein and Jordan frames and physicalness of these frames has always been controversial \cite{Kaiser1995, Kaiser1995a, Faraoni1999, Postma2014, Banerjee2016}. The conformal transformations define the induced degrees of freedom as scalar fields in the extended theories of gravity and consequently these transformations are used to investigate the models with different couplings between matter-energy content and the geometry. Indeed, the conformal transformations indicate the mathematical equivalence between the scalar-tensor gravity and the Einstein general relativity.

In the conformal transformations, the metric re-scaling which is dependent on the spacetime, is considered in the form
\begin{equation}‎\label{ct}
{g_{\mu \nu }} \rightarrow {\tilde g_{\mu \nu }} = {\Omega ^2}{g_{\mu \nu }},
‎\end{equation}
that we specify quantities in the Einstein frame by tilde. For the scalar-tensor gravities in which $ f(R,\phi ) =F(\phi)R$, the transformation parameter becomes
\begin{equation}\label{Omega}‎
\Omega^{2} =F\equiv \frac{\partial f}{\partial R}‎, ‎\hspace{.5cm} F>‎0.
\end{equation}
As a result, the action (\ref{Sst}) in the Einstein frame turns into
\begin{align}
\nonumber
{S_E} = \int d {x^4}\sqrt { - \tilde g} \left[ {\frac{1}{2}\tilde R - \frac{1}{2}{{\tilde g}^{\mu \nu }}{\partial _\mu }\phi {\partial _\nu }\phi  - V(\phi )} \right],
\label{SstE}
\end{align}
that now there exist no coupling between the Ricci scalar and the scalar field. In the above equation, $\phi$ and $V(\phi)$ are the scalar field and the potential in the Einstein frame, respectively. In order to the kinetic energy have the canonical form, we define the scalar field in the Einstein frame as
\begin{equation}‎\label{phivarphi}
\phi  = \int d \varphi \sqrt {\frac{3}{2}{{\left( {\frac{{{F_{,\varphi }}}}{F}} \right)}^2} + \frac{{\omega (\varphi )}}{F}}.
‎\end{equation}
Due to conformal transformation, the time and scale factor change as
\begin{equation}\label{taE}‎
‎d\tilde{t}=\sqrt{F}dt‎, ‎\hspace{.5cm} \tilde{a}=\sqrt{F}a‎.
‎\end{equation}
Therefore, the Hubble parameter changes in the form
\begin{equation}‎\label{HE}
\tilde H = \frac{1}{{\tilde a}}\frac{{d\tilde a}}{{d\tilde t}} = \frac{1}{{\sqrt F }}\left( {H + \frac{{\dot F}}{{2F}}} \right).
\end{equation}
In addition, the potential in the Einstein frame is given by \cite{Faraoni2004,Fujii2004}
\begin{equation}‎\label{VU}
V(\phi ) = {\left. {\frac{{U(\varphi )}}{{{F^2}(\varphi )}}} \right|_{\varphi  = \varphi (\phi )}}.‎
‎\end{equation}

If we have the scalar field and Hubble parameter in the Einstein frame, we can calculate the scalar power spectrum from \cite{Baumann2009}
\begin{equation}\label{PsE}
{\tilde {\cal P}_s} = {\left( {\frac{{\tilde H}}{{2\pi }}} \right)^2}{\left( {\frac{{\tilde H}}{{\phi '}}} \right)^2},
\end{equation}
where prime denotes derivative with respect to time in the Einstein frame. Also, if we obtain the potential in the Einstein frame, then we can simply calculate the observational parameters in terms of the potential slow-roll parameters which are expressed in terms of the potential and its derivatives as
\begin{align}
\label{epsV}
& {\varepsilon _V} \equiv \frac{1}{2}{\left( {\frac{{{V_{,\phi }}}}{V}} \right)^2},
\\
\label{etaV}
& {\eta _V} \equiv \frac{{{V_{,\phi \phi }}}}{V}.
\end{align}
In the Einstein frame, we can express the scalar spectral index and tensor-to-scalar ratio in terms of the potential slow-roll parameters as \cite{Baumann2009}
\begin{align}
\label{nsE}
& {\tilde{n}_s}  \simeq  1 + 2{\eta _V} - 6{\varepsilon _V},
\\
\label{rE}
& \tilde{r} \simeq 16 \varepsilon_{V},
\end{align}
which are valid in the slow-roll approximation.

\section{Inflation in the Brans-Dicke gravity}\label{sec:BD}

In this section, we consider the Brans-Dicke gravity as a special model of the scalar-tensor gravity and derive the background field equations in this model. Then, we turn to study inflation in this model and using the relations expressed in the previous section, we obtain the observational quantities for the Brans-Dicke gravity in both the Jordan and Einstein frames. In the next section, we will use the results for the inflationary observables for different potentials which have motivations from quantum field theory or string theory. In this way, we will be able to compare behaviors of those potentials in the Brans-Dicke gravity versus their behaviors in the standard inflationary scenario based on the Einstein gravity. Furthermore, we will check viability of those inflationary potentials in light of the Planck 2015 observational data.

Brans and Dicke \cite{Brans1961} proposed a specific form of the scalar-tensor gravity that it is founded on the Mach principle, which implies that the inertial mass of an object depends on the matter distribution in the universe and thus the gravitational constant should have time-dependence. This idea was in agreement with Dirac's prediction about the time-dependence of the gravitational constant so that the quantities constructed from the fundamental constants, take the values of order of the elementary particles. In the Brans-Dicke theory a scalar field is invoked to describe the time-dependence of the gravitational constant. In order to the action (\ref{Sst}) turn into the action of the Brans-Dicke gravity, we should consider
\begin{equation}\label{BD}‎
‎f(R,\varphi)=\varphi R‎ , ‎\hspace{.5cm} \omega(\varphi)=\frac{\omega_{BD}}{\varphi}‎,
‎\end{equation}
where $\omega_{BD}$ is the Brans-Dicke parameter which is a constant. Therefore, the form of the Brans-Dicke action in the Jordan frame becomes
\begin{equation}‎\label{SBD}
S_{J} = \int {{d^4}} x\sqrt { - g} \left[ {\frac{1}{2}\varphi R - \frac{1}{2}\frac{{{\omega _{BD}}}}{\varphi }{g^{\mu \nu }}{\partial _\mu }\varphi {\partial _\nu }\varphi  - U(\varphi )} \right].
‎\end{equation}
Hereafter, we drop out the subscript ``BD'' in the Brans-Dicke parameter and write it as $\omega$. It should be noted that the original Brans-Dicke theory does not contain the potential ($U(\varphi ) = 0$) \cite{Brans1961}.

Using Eqs. (\ref{3HF}) and (\ref{varphiddot}) for the Brans-Dicke action (\ref{SBD}), we obtain the evolution equations for a spatially flat FRW universe as
\begin{align}
\label{HU}
& 3{\left( {H + \frac{{\dot \varphi }}{{2\varphi }}} \right)^2} - \frac{{\left( {2\omega  + 3} \right)}}{4}{\left( {\frac{{\dot \varphi }}{\varphi }} \right)^2}-\frac{U}{\varphi } = 0,
\\
\label{varphiddotBDwp}
& \ddot \varphi  + 3H\dot \varphi  + \frac{2}{{\left( {2\omega  + 3} \right)}}\left( {\varphi {U_{,\varphi }} - 2U} \right) = 0.
\end{align}
Considering the slow-roll conditions $\left| {\dot \varphi } \right| \ll \left| {H\varphi } \right|$ and $\left| {\ddot \varphi } \right| \ll \left| {3H\dot \varphi } \right|$, Eqs. (\ref{HU}) and (\ref{varphiddotBDwp}) reduce to
\begin{align}
\label{HUsr}
& 3{H^2}\varphi  - U \simeq 0,
\\
\label{varphidotUsr}
& 3H\dot \varphi  + \frac{2}{{\left( {2\omega  + 3} \right)}}\left( {\varphi {U_{,\varphi }} - 2U} \right) \simeq 0.
\end{align}
From Eqs. (\ref{HUsr}) and (\ref{varphidotUsr}), one can get $H$ and $\dot{\varphi}$ in terms of the potential $U(\varphi)$ in the slow-roll approximation.

Here, we introduce the $e$-fold number which is used to determine the amount of inflation and is defined as
\begin{equation}
\label{N}
N \equiv \ln \left({\frac{{{a_e}}}{a}} \right),
\end{equation}
where $a_e$ is the scale factor at the end of inflation. The above definition gives rise to
\begin{equation}
\label{dN}
dN =  - H dt = - \frac{H}{{\dot \varphi }}d\varphi.
\end{equation}
The anisotropies observed in the CMB correspond to the perturbations whose wavelengths crossed the Hubble radius around $N_* \approx 50 - 60$ before the end of inflation \cite{Liddle2003, Dodelson2003}. This result can be obtained with the assumption that during inflationary era, a slow-roll inflation has occurred that it provides a quasi-de Sitter expansion with $H \approx {\rm{constant}}$ for the universe. In addition, the evolution of the universe after inflation is assumed to be determined by the standard model of cosmology. In this work, we have used these two assumptions and thus we can take the $e$-folds number of the horizon crossing as $N_* \approx 50 - 60$ from the end of inflation. Substituting $H$ and $\dot{\varphi}$ from Eqs. (\ref{HUsr}) and (\ref{varphidotUsr}), respectively, into Eq. (\ref{dN}), we obtain
\begin{equation}\label{Nsr}
N \simeq \frac{{\left( {2\omega  + 3} \right)}}{2}\int_{{\varphi _e}}^\varphi  {\frac{U}{{\varphi \left( {\varphi {U_{,\varphi }} - 2U} \right)}}} d\varphi,
\end{equation}
where $\varphi_e$ is the scalar field at the end of inflation that to determine it, we use the relation ${\varepsilon _1} = 1$, because the slow-roll conditions are violated at the end of inflation.

From Eqs. (\ref{E}) and (\ref{BD}), we see that the parameter $E$ for the Brans-Dicke gravity becomes a constant as $E = \omega  + 3/2$, and therefore the fourth slow-roll parameter in Eq. (\ref{eps}) vanishes ($\varepsilon_{4}=0$). Consequently, the scalar spectral index for this model results from Eq. (\ref{nseps}) as
\begin{equation}\label{nsBD}
n{_s} \simeq 1 - 4{\varepsilon _1} - 2{\varepsilon _2} + 2{\varepsilon _3}.
\end{equation}
From the above equation, we calculate the running of the scalar spectral index for the Brans-Dicke gravity as
\begin{equation}\label{dnsBD}
\frac{{d{n_s}}}{{d\ln k}} \simeq  - 8\varepsilon _1^2 + 2\varepsilon _2^2 - 4\varepsilon _3^2 - 2{\varepsilon _1}{\varepsilon _2} + 4{\varepsilon _1}{\varepsilon _3},
\end{equation}
that we have used the relation $k=a H$ which is valid at the horizon crossing. Within the framework of Brans-Dicke gravity, we get the parameter $Q_s$ from Eq. (\ref{Qs2}) as
\begin{equation}\label{QsBD}
{Q_s} = \frac{{{{\dot \varphi }^2}\left( {2\omega  + 3} \right)}}{{{2H^2}\varphi {{\left( {1 + \frac{{\dot \varphi }}{{2H\varphi }}} \right)}^2}}}.
\end{equation}
Substituting the above result into Eq. (\ref{rst}), we obtain the tensor-to-scalar ratio for the Brans-Dicke gravity as
\begin{equation}\label{rBD}
r \simeq 4\left( {2\omega  + 3} \right)\frac{{{{\dot \varphi }^2}}}{{{H^2}{\varphi ^2}}}.
\end{equation}

In the following, we try to find the inflationary observables in terms of the potential. To this aim, it is useful to find expressions of the slow-roll parameters in the slow-roll approximation. If we use Eqs. (\ref{HUsr}) and (\ref{varphidotUsr}) in (\ref{eps}), we get the non-vanishing slow-roll parameters
\begin{align}
\label{eps1}
& {\varepsilon _1} = \frac{{\left( {\varphi {U_{,\varphi }} - 2U} \right)\left( {\varphi {U_{,\varphi }} - U} \right)}}{{\left( {2\omega  + 3} \right){U^2}}},
\\
\label{eps2}
& {\varepsilon _2} = {\varepsilon _1} - \frac{{2\varphi \left( {\varphi {U_{,\varphi \varphi }} - {U_\varphi }} \right)}}{{\left( {2\omega  + 3} \right)U}},
\\
\label{eps3}
& {\varepsilon _3} =  - \frac{{\left( {\varphi {U_{,\varphi }} - 2U} \right)}}{{\left( {2\omega  + 3} \right)U}}.
\end{align}
Consequently, if we use Eqs. (\ref{HUsr}) and (\ref{varphidotUsr}) in Eq. (\ref{QsBD}) and then insert the result into Eq. (\ref{PsH}), we obtain the scalar power spectrum in terms of the inflationary potential as
\begin{equation}\label{PsU}
{{\cal P}_s} \simeq \frac{{\left( {2\omega  + 3} \right){U^3}}}{{24{\pi ^2}{\varphi ^2}{{\left( {\varphi {U_{,\varphi }} - 2U} \right)}^2}}}.
\end{equation}
In addition, substituting the slow-roll parameters (\ref{eps1}), (\ref{eps2}) and (\ref{eps3}) into Eq. (\ref{nsBD}), the scalar spectral index takes the form
\begin{equation}\label{nsU}
{n_s} \simeq 1 + \frac{2}{{\left( {2\omega  + 3} \right){U^2}}}\left[ {\varphi \left( {6U{U_{,\varphi }} + 2\varphi U{U_{,\varphi \varphi }} - 3\varphi U_\varphi ^2} \right) - 4{U^2}} \right].
\end{equation}
Moreover, using Eqs. (\ref{HUsr}) and (\ref{varphidotUsr}), the tensor-to-scalar ratio (\ref{rBD}) is obtained as
\begin{equation}\label{rU}
r \simeq \frac{{16{{\left( {\varphi {U_{,\varphi }} - 2U} \right)}^2}}}{{\left( {2\omega  + 3} \right){U^2}}}.
\end{equation}

Another inflationary observable which can be used to discriminate between inflationary models, is the non-Gaussianity parameter (for review see e.g. \cite{Bartolo2004, Chen2010}). Different inflationary models predict maximal signal for different shapes of non-Gaussianity. Therefore, the shape of non-Gaussianity is potentially a powerful probe of the mechanism that generate the primordial perturbations \cite{Babich2004, Baumann2009}. For the single field inflationary models with non-canonical kinetic terms, the non-Gaussianity parameter has peak in the equilateral shape. Also, the squeezed shape is the dominant mode of models with multiple light fields during inflation. Furthermore, the folded non-Gaussianity becomes dominant in models with non-standard initial states.

The subject of primordial non-Gaussianities in the Brans-Dicke theory has been investigated in details in \cite{Felice2011}. However, since in the present work, we deal with a single field inflation with a non-canonical kinetic term and standard initial states (such as the Bunch-Davies vacuum initial conditions for perturbations), therefore we focus on the non-Gaussianity parameter in the equilateral limit. The equilateral non-Gaussianity parameter for the Brans-Dicke gravity has been obtained in \cite{Artymowski2015} as
\begin{equation}\label{fNLequil}
‎f_{{\rm{NL}}}^{{\rm{equil}}} =  - \frac{5}{4}{\varepsilon _2} + \frac{5}{6}{\varepsilon _3}.
\end{equation}
We see that in the Brans-Dicke gravity, the equilateral non-Gaussianity is of order of the slow-roll parameters which are very smaller than unity in the slow-roll regime. On the other hand, the slow-roll conditions can be perfectly satisfied in the Brans-Dicke gravity. Therefore, the equilateral non-Gaussianity parameter in the Brans-Dicke gravity can be in agreement with the Planck 2015 prediction, $f_{{\rm{NL}}}^{{\rm{equil}}} =  - 16 \pm 70$ (68\% CL, Planck 2015 T-only), see \cite{Planck2015}. We will show this fact in the next section explicitly for different inflationary potentials.

At the end of this section, we discus about equivalence of the results for the inflationary observables in the Jordan and Einstein frames. We saw before that via the conformal transformation $\tilde{g}=\Omega^{2}g$, we can go from the Jordan frame to the Einstein frame. For the Brans-Dicke gravity, $F=\varphi$, and thus from Eq. (\ref{taE}) we see that the time and scale factor change under the conformal transformation as
\begin{equation}\label{taEBD}
d\tilde t = \sqrt \varphi~  dt‎, ‎\hspace{.5cm} \tilde a = \sqrt \varphi~  a‎.
\end{equation}
Also, from Eq. (\ref{HE}) we conclude that the Hubble parameter transforms in the form
\begin{equation}\label{HEBD}
\tilde H = \frac{H}{{\sqrt \varphi  }}.
\end{equation}
To find the relation between the scalar fields in the Einstein and Jordan frames in the Brans-Dicke gravity, we use Eq. (\ref{phivarphi}) and get
\begin{equation}\label{phivarphiBD}
\phi  = \sqrt {\frac{{2\omega  + 3}}{2}} \ln \varphi‎.
\end{equation}
Also, from Eq. (\ref{VU}), we see that the relation between potentials in the two frames is
\begin{equation}\label{VUBD}
V(\phi ) = {\left. {\frac{{U(\varphi )}}{{{\varphi ^2}}}} \right|_{\varphi  = \varphi (\phi )}}.
\end{equation}

Here, we want to know how the inflationary observables change under the conformal transformation from the Jordan frame to the Einstein frame in the Brans-Dicke gravity. First, we focus on the transformation of the scalar power spectrum. If we use Eqs. (\ref{taEBD}), (\ref{HEBD}) and (\ref{phivarphiBD}) in Eq. (\ref{PsE}), and compare the result with Eq. (\ref{PsU}), we conclude that
\begin{equation}\label{PsEJ}
{\tilde {\cal P}_s} \simeq {{\cal P}_s},
\end{equation}
which implies that in the slow-roll approximation, the equations for the scalar power spectrum are same in both the Einstein and Jordan frames.

In what follows, we proceed to find the transformations of the scalar spectral index and tensor-to-scalar ratio. To do so, we can use Eqs. (\ref{phivarphiBD}) and (\ref{VUBD}) in Eqs. (\ref{epsV}) and (\ref{etaV}), and obtain the potential slow-roll parameters in the Einstein frame in terms of the scalar field $\varphi$ and potential $U(\varphi)$ in the Jordan frame, as
\begin{align}
\label{epsVBD}
& {\varepsilon _V} = \frac{{{{\left( {\varphi {U_\varphi } - 2U} \right)}^2}}}{{\left( {2\omega  + 3} \right){U^2}}},
\\
\label{etaVBD}
& {\eta _V} = {\varepsilon _V} + \frac{{2\varphi \left( {{U_{,\varphi \varphi }} - {U_\varphi }} \right)}}{{\left( {2\omega  + 3} \right)U}} - \frac{{{\varphi ^2}U_{,\varphi }^2}}{{\left( {2\omega  + 3} \right){U^2}}} + \frac{4}{{\left( {2\omega  + 3} \right)}}.
\end{align}
If we use these relations in Eqs. (\ref{nsE}) and (\ref{rE}), and compare the result with Eqs. (\ref{nsU}) and (\ref{rU}), then we see that
\begin{align}
\label{nsEJ}
& \tilde{n}_{s} \simeq n_{s}‎,
\\
\label{rEJ}
‎& \tilde{r} \simeq r,
\end{align}
which means that in the Brans-Dicke gravity and in the slow-roll approximation,
the relations for the scalar spectral index and tenor-to-scalar ratio are identical in the Einstein and Jordan frames.

There are much discussion and challenge about the Jordan and Einstein frames and also about the results corresponding to the inflationary observables in these two frames \cite{Kaiser1995, Kaiser1995a, Faraoni1999,  Postma2014, Banerjee2016, Makino1991, Fakir1990, Salopek1989, Tsujikawa2004, Hinterbichler2012, Chiba2008, Gong2011, Catena2007, Chiba2013}. In fact, the conformal invariance of the scalar power spectrum from the Jordan frame to the Einstein frame, Eq. (\ref{PsEJ}), is expected because of the conformal invariance of the curvature perturbation, $\tilde {\cal R} = {\cal R}$ (for more details, see \cite{Felice2010}). The conformal invariance of the amplitude of scalar perturbations was firstly shown in \cite{Makino1991}, for $\lambda {\varphi ^4}$ chaotic inflation model with the non-minimal coupling $\xi {\varphi ^2}R$ in a more rigorous manner relative to the previous papers \cite{Fakir1990, Salopek1989}. In \cite{Tsujikawa2004}, the authors have generalized the conformal invariance of both the scalar and tensor power spectra for the model with non-minimal coupling term $F(\varphi)R$. In \cite{Kaiser1995}, it has been clarified that the scalar spectral index for the new inflation model \cite{Linde1982, Albrecht1982}, is different in the two frames, but for the chaotic inflation model \cite{Linde1983} with various initial conditions, the results are identical in both frames. Furthermore, in \cite{Kaiser1995a}, it has been discussed that if one applies the slow-roll approximation in obtaining the scalar power spectrum, then the scalar spectral index for both the new and chaotic inflation models with various initial conditions, are same in the two frames. In \cite{Felice2010}, it was shown that the curvature perturbation and the tensor perturbations remain invariant under the conformal transformations, and hence the scalar and tensor power spectrum remain invariant in the two frames. Consequently, the tensor-to-scalar ratio is identical for the both frames. In the present paper, our study implies that for the Brans-Dicke gravity, the relations of the scalar power spectrum ${\cal P}_s$, the scalar spectral index ${n_s}$  and the tensor-to-scalar ratio $r$ are same in the two frames, only in the slow-roll approximation. It is worth mentioning that the conformal invariance holds for the adiabatic modes even beyond the slow-roll approximation \cite{Hinterbichler2012}. The conformal equivalence of the inflationary observables also holds at the non-linear level as shown in \cite{Chiba2008, Gong2011}. Furthermore, it was pointed out that the other cosmological observables/relations besides the inflationary ones, such as redshift, luminosity distance, temperature anisotropies, cross sections, etc. are frame-independent \cite{Catena2007, Chiba2013}.

\section{Study of various inflationary potentials in the Brans-Dicke gravity}\label{sec:pot}

In the previous section, we obtained the relations of the inflationary observables in the Brans-Dicke gravity using the Jordan frame. Here, we apply the results of the previous section for various inflationary potentials and check their viability in light of the Planck 2015 observational results \cite{Planck2015}. We examine the potentials which have motivations from quantum field theory or string theory. Note that validity of these potentials in comparison with the observational data has been already investigated in \cite{Martin2014a, Martin2014b, Rezazadeh2015, Huang2015, Okada2014}, but within the framework of standard inflationary scenario based on Einstein's gravity.

To examine each potential, first we use it in Eqs. (\ref{nsU}) and (\ref{rU}), and find the scalar spectral index $n_s$ and the tensor-to-scalar ratio $r$ in terms of the inflaton scalar field $\varphi$. Then, we set ${\varepsilon _1} = 1$ in Eq. (\ref{eps1}) to determine analytically the scalar field at the end of inflation, ${\varphi _e}$. Next, we use ${\varphi _e}$ in Eq. (\ref{Nsr}) and apply a numerical method to obtain the inflaton scalar field at the horizon exit, $\varphi_*$, that we take the $e$-fold number of the epoch of horizon exit as ${N_*} = 50$ or $60$. In this way, we can evaluate $n_s$ and $r$ at the horizon exit and then plot the $r-n_s$ diagram for the model. Finally, comparing the result of the model in $r-n_s$ plane with the allowed region by the Planck 2015 data \cite{Planck2015}, we are able to check viability of the considered inflationary potential in light of the observational results.

\subsection{Power-law potential}\label{subsec:pl}

We start with the simplest inflationary potential which is the power-law potential
\begin{equation}\label{Upl}
U(\varphi ) = {U_0}{\varphi ^n},
\end{equation}
where $U_0$ and $n>0$ are constant parameters. This class of potentials includes the simplest chaotic inflationary models introduced by \cite{Linde1983}, in which inflation starts from large values for the inflaton, i.e. $\varphi > M_P$. In the standard inflationary scenario, this potential can be in agreement with Planck 2015 TT,TE,EE+lowP data \cite{Planck2015} at 95\% CL, as it has been shown in \cite{Rezazadeh2015}.

In the Brans-Dicke gravity setting, the power-law potential (\ref{Upl}) with $n>2$ leads to the power-law inflation with the scale factor $a(t) \propto t^q$ where $q>1$ \cite{Artymowski2015}. Therefore, the slow-roll parameters (\ref{eps}) turn to be constant and they become dependent on the parameters $n$ and $\omega$. As a result, the scalar spectral index $n_s$ and the tensor-to-scalar ratio $r$ become constant as
\begin{align}
\label{nspl}
& {n_s} = 1 - \frac{{2{{\left( {n - 2} \right)}^2}}}{{2\omega  + 3}},
\\
\label{rpl}
& r = \frac{{16{{\left( {n - 2} \right)}^2}}}{{2\omega  + 3}}.
\end{align}

\begin{figure}[t]
\begin{center}
\scalebox{1}[1]{\includegraphics{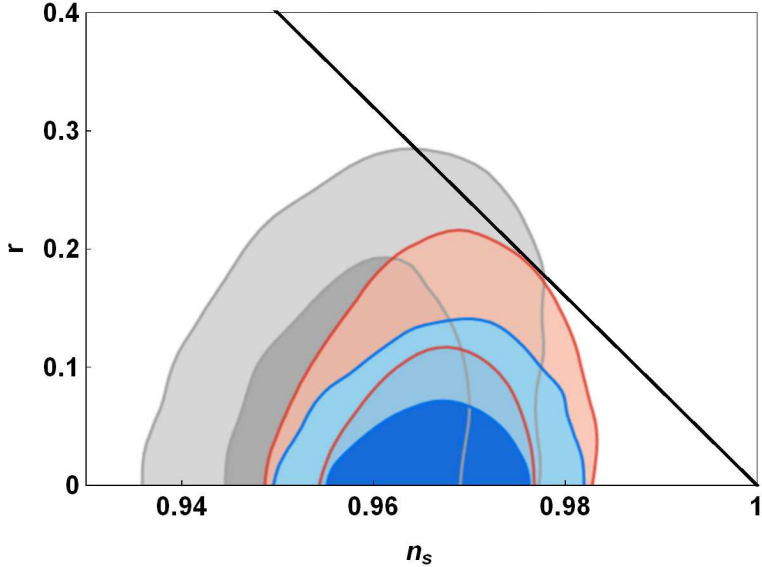}}
\caption{Prediction of power-law potential (\ref{Upl}) in $r-n_s$ plane in the Brans-Dicke gravity (black line). The marginalized joint 68\% and 95\% CL regions of Planck 2013, Planck 2015 TT+lowP and Planck 2015 TT,TE,EE+lowP data \cite{Planck2015} are specified by gray, red and blue, respectively.}
\label{fig:pl}
\end{center}
\end{figure}

From the two above equations, we see that for large values of the Brans-Dicke parameter $\omega$, the scalar spectral index $n_s$ approaches unity while the tensor-to-scalar ratio $r$ converges to zero. We see that the two above equations can be easily combined to give the linear relation
\begin{equation}\label{rnspl}
r = 8\left( {1 - {n_s}} \right).
\end{equation}
This relation implies that the prediction of the power-law potential (\ref{Upl}) in $r-n_s$ plane is independent of the parameters $\omega$ and $n$. Using the above relation, we can draw the $r-n_s$ plot as shown by a black line in Fig. \ref{fig:pl}. Moreover, in Fig. \ref{fig:pl}, the marginalized joint 68\% and 95\% confidence limit (CL) regions for Planck 2013, Planck 2015 TT+lowP and Planck 2015 TT,TE,EE+lowP data \cite{Planck2015} are specified by gray, red and blue, respectively. The figure shows that (i) the result of the power-law potential in the Brans-Dicke gravity in contrary to the standard model, lies outside the range allowed by the Planck 2015 data. (ii) The prediction of this potential takes place in the region 95\% CL of Planck 2013 data. This is in good agreement with that obtained by \cite{Felice2011a} using the data of WMAP7.

\subsection{Inverse power-law potential}\label{subsec:ipl}

The next potential which we examine is the inverse power-law potential
\begin{equation}\label{Uipl}
U(\varphi ) = {U_0}{\varphi ^{ - n}},
\end{equation}
where $U_0$ and $n>0$ are two model parameters. This potential is a steep potential and in the standard inflationary setting, it gives rise to the intermediate inflation with the scale factor $a(t) \propto \exp [A{({M_P}t)^\lambda}]$ where $A>0$ and $0<\lambda<1$ \cite{Barrow1990, Barrow2006, Barrow2007}, which is not consistent with the Planck 2015 observational results, as it has been discussed in \cite{Rezazadeh2015}.

To obtain the equations of $n_s$ and $r$ for this potential in the Brans-Dicke theory, we can simply change $n \to  - n$ in Eqs. (\ref{nspl}) and (\ref{rpl}). In this way, if we can combine the results, we again recover relation (\ref{rnspl}) between $n_s$ and $r$. Therefore, the $r-n_s$ plot for the inverse power-law potential (\ref{Uipl}) becomes like the one for the power-law potential (\ref{Upl}) which it has been shown in Fig. \ref{fig:pl}. Consequently, inflation with the inverse power-law potential in the Brans-Dicke gravity like the standard setting is ruled out by the Planck 2015 data.

\subsection{Exponential potential}\label{subsec:exp}

Another steep potential that we study, is the exponential potential
\begin{equation}\label{Uexp}
U(\varphi ) = {U_0}{e^{ - \alpha \varphi }},
\end{equation}
where $U_0$ and $\alpha >0$ are constant parameters. In the standard inflation model, this potential provides the power-law inflation with the scale factor $a(t) \propto t^q$ where $q>1$ \cite{Lucchin1985, Halliwell1987, Yokoyama1988}, that cannot be compatible with the Planck 2015 results, as it has been demonstrated in \cite{Rezazadeh2015, Rezazadeh2016}.

Within the framework of Brans-Dicke gravity, the observables $n_s$ and $r$ for the exponential potential (\ref{Uexp}) become independent of the parameters $U_0$ and $\alpha$. We can evaluate $n_s$ and $r$ for different values of the Brans-Dicke parameter $\omega$ and the horizon exit $e$-fold number $N_*$. Our examination shows that with $N_*=50$ and $N_*=60$, the tensor-to-scalar ratio for different values of $\omega$, varies in the ranges $r \ge 0.687$ and $r \ge 0.580$, respectively. These results for $r$ are not consistent with the upper bound $r< 0.149$ (95\% CL) deduced from Planck 2015 TT,TE,EE+lowP data \cite{Planck2015}. Therefore, the exponential potential (\ref{Uexp}) in the Brans-Dicke gravity like the standard scenario is disfavored by the observational data.

\subsection{Hilltop potential}\label{subsec:hil}

A potential which has a remarkable importance in study of inflation is the hilltop potential
\begin{equation}\label{Uhil}
U(\varphi ) = {U_0}\left( {1 - \frac{{{\varphi ^p}}}{{{\mu ^p}}} + ...} \right),
\end{equation}
where $U_0$, $\mu$ and $p>0$ are constant parameters of the model \cite{Boubekeur2005}. In this interesting class of potentials, the inflaton rolls away from an unstable equilibrium as in the first new inflationary models \cite{Linde1982, Albrecht1982}. This potential in the standard inflationary scenario can be in excellent agreement with the Planck 2015 results, because its prediction can lie inside the region 68\% CL of Planck 2015 TT,TE,EE+lowP data \cite{Planck2015}.

\begin{figure}[t]
\begin{center}
\scalebox{1}[1]{\includegraphics{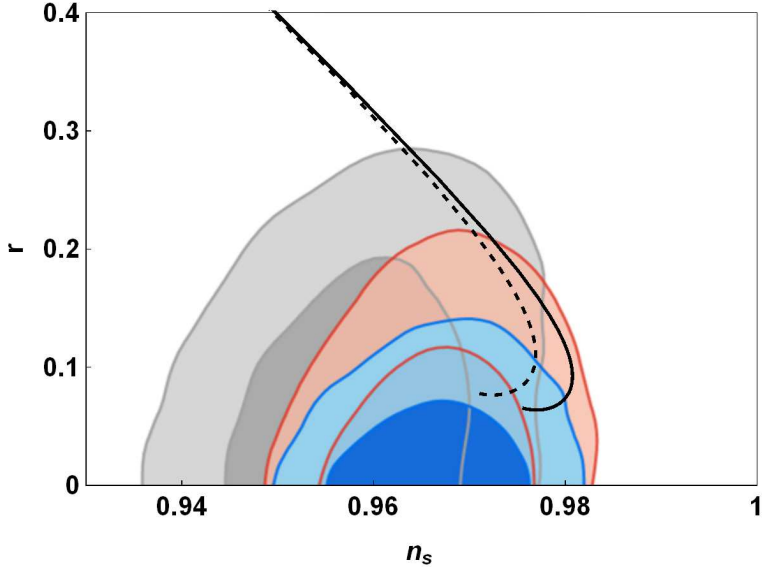}}
\caption{Same as Fig. \ref{fig:pl} but for the hilltop potential (\ref{Uhil}). The result of the potential for $N_*=50$ and $N_*=60$ are shown by the dashed and solid black lines, respectively.}
\label{fig:hil}
\end{center}
\end{figure}

Study of this potential in the Brans-Dicke gravity shows that the quantities $n_s$ and $r$ do not depend on $U_0$ and $\mu$. Furthermore, we conclude that for $p = 1,\,2,\,3,\,4,\,6$, results of this potential can be placed inside the region 95\% CL of Planck 2015 TT,TE,EE+lowP data, if we increase the parameter $\omega$ sufficiently. In Fig. \ref{fig:hil}, the $r-n_s$ plot for the hilltop potential (\ref{Uhil}) with $p=4$ is illustrated in comparison with the observational data. In the figure, the results of the model with $N_*=50$ and $N_*=60$ are shown by the dashed and solid black lines, respectively. Here, it is worthwhile to mention that in this figure, the result of the model lies inside the region 95\% CL for $\omega \gtrsim {10^3}$. In this model, a large values of the parameter $\omega$ gives larger values for $n_s$ and $r$. For very large values of $\omega$ relative to unity, the observables $n_s$ and $r$ approach $0.9708$ ($0.9757$) and $0.08$ ($0.07$), respectively, if we take the $e$-fold number of horizon crossing as $N_*=50$ ($N_*=60$).

\subsection{D-brane potential}\label{subsec:Db}

Another inflationary potential which has motivations from the physical theories with extra dimensions is the D-brane potential
\begin{equation}\label{UDb}
U(\varphi ) = {U_0}\left( {1 - \frac{{{\mu ^p}}}{{{\varphi ^p}}} + ...} \right),
\end{equation}
where $U_0$, $\mu$ and $p>0$ are constant parameters. Two important cases of this potential correspond to $p=2$ \cite{Garcia-Bellido2002} and $p=4$ \cite{Dvali2001,Kachru2003} are compatible with the Planck 2015 data in the standard inflationary scenario, as mentioned in \cite{Planck2015}.

\begin{figure}[t]
\begin{center}
\scalebox{1}[1]{\includegraphics{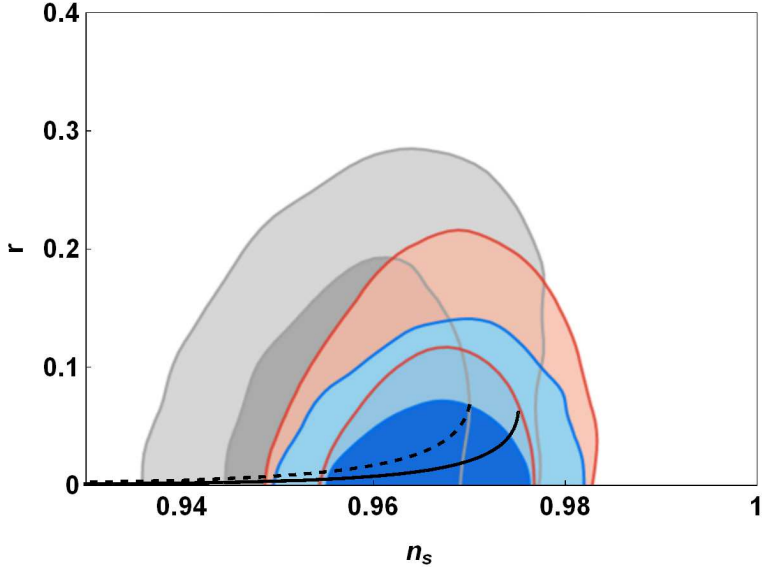}}
\caption{Same as Fig. \ref{fig:pl} but for the D-brane potential (\ref{UDb}). The results for $N_*=50$ and $N_*=60$ are shown by the dashed and solid black lines, respectively.}
\label{fig:Db}
\end{center}
\end{figure}

In the inflationary scenario based on the Brans-Dicke gravity, the observables $n_s$ and $r$ for the D-brane potential (\ref{UDb}) depend only on $\omega$ and $N_*$. For $N_*=50$ and $N_*=60$, the results of this potential with $p=2,\, 4$ lie inside the 68\% CL region of Planck 2015 TT,TE,EE+lowP data. We see this fact for the case $p=4$ in Fig. \ref{fig:Db}, that the dashed and solid black lines correspond to $N_*=50$ and $N_*=60$, respectively. In this figure, as the parameter $\omega$ increases, the observables $n_s$ and $r$ grow and converge respectively to $0.9701$ ($0.9751$) and $0.08$ ($0.07$), for the $e$-fold number of horizon exit $N_*=50$ ($N_*=60$).

\subsection{Higgs potential}\label{subsec:Hig}

Now, we investigate the Higgs potential
\begin{equation}\label{UHig}
U(\varphi ) = {U_0}{\left[ {1 - {{\left( {\frac{\varphi }{\mu }} \right)}^2}} \right]^2},
\end{equation}
where $U_0$ and $\mu>0$ are model parameters \cite{Kallosh2007, Okada2014}. This potential leads to a mechanism of symmetry breaking, where the field rolls off an unstable equilibrium toward a displaced vacuum \cite{Baumann2009}. The Higgs potential (\ref{UHig}) in the standard inflationary scenario behaves like a small-field potential when $\varphi<\mu$, and like a large-field potential when $\varphi>\mu$. In \cite{Rezazadeh2015} it has been shown that the result of this potential in the standard inflationary framework can lie within the joint 68\% CL region of Planck 2015 TT,TE,EE+lowP data.

Within the framework of Brans-Dicke gravity, our results for the Higgs potential (\ref{UHig}) show that the observables $n_s$ and $r$ does not depend on the parameters $U_0$ and $\mu$. Surprisingly, we found that the result of this potential for $n_s$ and $r$ are completely identical for both sides of the potential minimum, i.e. for $\varphi<\mu$ and $\varphi>\mu$. In order to explain this unexpected result, we note that to know whether the result of a potential in $r-n_s$ plane is the same for the both sides of its minimum in the Jordan frame, we should go to the Einstein frame via the conformal transformation. If the shape of the potential is symmetric around its minimum in the Einstein frame, then its prediction in $r-n_s$ plane will be the same for the both sides of its minimum, and otherwise the results will be different. For the Higgs potential (\ref{UHig}), using Eqs. (\ref{phivarphiBD}) and (\ref{VUBD}), it changes under the conformal transformation as
\begin{equation}\label{VHig}
V(\phi ) = \frac{{{U_0}}}{\mu }{\left( {{e^{\sqrt {\frac{2}{{2\omega  + 3}}}~\phi }} - {\mu ^2}{e^{ - \sqrt {\frac{2}{{2\omega  + 3}}}~\phi }}} \right)^2}.
\end{equation}
The above potential is completely symmetric around its minimum, $\tilde \mu  = \sqrt {\frac{{2\omega  + 3}}{2}} \ln \mu $. Consequently, although the Higgs potential (\ref{UHig}) is not symmetric around its minimum in the Jordan frame, the transformed potential (\ref{VHig}) is completely symmetric around its minimum in the Einstein frame. In addition, we should note that the values of $\phi  < \tilde \mu $ and $\phi  > \tilde \mu $ in the Einstein frame are related respectively to the values $\varphi  < \mu $ and $\varphi  > \mu $, in the Jordan frame. As a result, we conclude that the results of Higgs potential (\ref{UHig}) for $n_s$ and $r$ should be same for the both regions $\varphi<\mu$ and $\varphi>\mu$.

\begin{figure}[t]
\begin{center}
\scalebox{1}[1]{\includegraphics{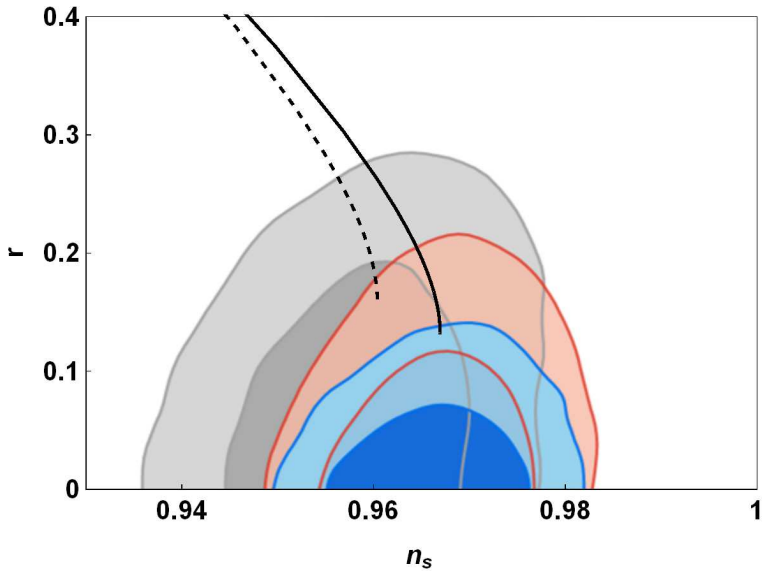}}
\caption{Same as Fig. \ref{fig:pl} but for the Higgs potential (\ref{UHig}). The results for $N_*=50$ and $N_*=60$ are shown by the dashed and solid black lines, respectively.}
\label{fig:Hig}
\end{center}
\end{figure}

The $r-n_s$ diagram for this potential is shown in Fig. \ref{fig:Hig}, and as we see in the figure, the result of the Higgs potential (\ref{UHig}) for $N_*=60$ can be placed inside the 95\% CL region of Planck 2015 TT,TE,EE+lowP data. For this model, as $\omega$ becomes larger, the scalar spectral index $n_s$ increases and approaches $0.9604$ ($0.9669$) for $N_*=50$ ($N_*=60$). Moreover, the greater $\omega$ leads to the smaller values for the tensor-to-scalar ratio $r$ and it finally approaches $0.16$ ($0.13$) for $N_*=50$ ($N_*=60$).

\subsection{Coleman-Weinberg potential}\label{subsec:CW}

A famous inflationary potential which has ideas from quantum field theory, is the Coleman-Weinberg potential
\begin{equation}\label{UCW}
U(\varphi ) = {U_0}\left[ {{{\left( {\frac{\varphi }{\mu }} \right)}^4}\left( {\ln \left( {\frac{\varphi }{\mu }} \right) - \frac{1}{4}} \right) + \frac{1}{4}} \right],
\end{equation}
with constants $U_0$ and $\mu>0$ \cite{Martin2014a, Martin2014b, Okada2014, Barenboim2014}. This potential is historically famous since it was applied in the original papers of the new inflation model \cite{Linde1982, Albrecht1982}. In \cite{Rezazadeh2015}, the authors have examined this potential in the standard inflationary framework and shown that it can be consistent with 68\% CL region of Planck 2015 TT,TE,EE+lowP data.

\begin{figure}[t]
\begin{center}
\scalebox{1}[1]{\includegraphics{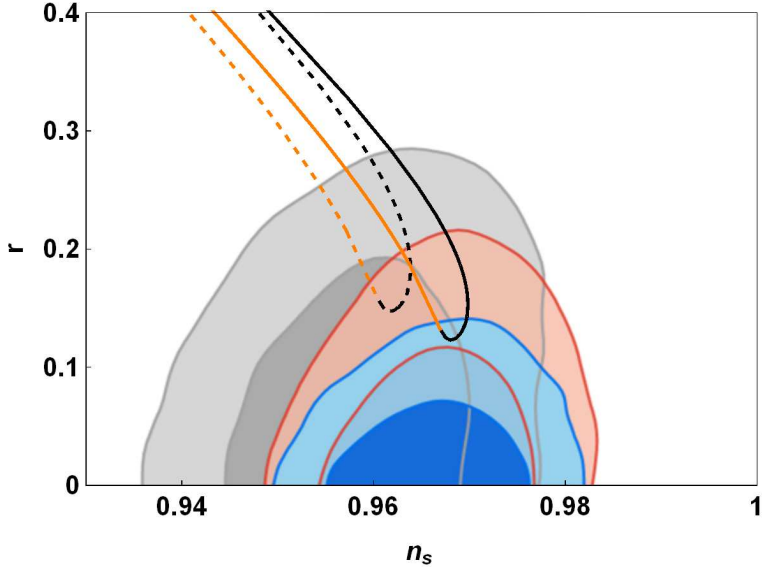}}
\caption{Same as Fig. \ref{fig:pl} but for the Coleman-Weinberg potential (\ref{UCW}). The predictions of the model for $N_*=50$ and $N_*=60$ are shown by the dashed and solid black lines, respectively. In addition, the results of this potential for the two ranges $\varphi<\mu$ and $\varphi>\mu$ are specified by black and orange colors, respectively.}
\label{fig:CW}
\end{center}
\end{figure}

The result of Coleman-Weinberg potential (\ref{UCW}) in the Brans-Dicke gravity is plotted in Fig. \ref{fig:CW}. In contrast with the Higgs potential (\ref{UHig}), the prediction of this potential for the two regimes $\varphi<\mu$ and $\varphi>\mu$ are completely different as shown in Fig. \ref{fig:CW} by black and orange colors, respectively. For the values $\omega  \gg 1$, the results of the two regimes approach to a common point in $r-n_s$ plane, that we see this behavior in Fig. \ref{fig:CW}. Also, from the figure we see that the prediction of the Coleman-Weinberg potential (\ref{UCW}) for the both regimes $\varphi<\mu$ and $\varphi>\mu$ can place within the joint 95\% CL region of Planck 2015 TT,TE,EE+lowP data, if we take the horizon exit $e$-fold number as $N_*=60$. For this potential, for the both cases $\varphi<\mu$ and $\varphi>\mu$, for large values of $\omega$, $n_s$ and $r$ converge respectively to $0.9604$ ($0.9669$) and $0.16$ ($0.13$), if we take the $e$-fold number of horizon exit as $N_*=50$ ($N_*=60$).

\subsection{Natural inflation}\label{subsec:nat}

In what follows, we concentrate on one of the most elegant inflationary models which is natural inflation given by the periodic potential \cite{Freese1990, Adams1993, Freese2014}
\begin{equation}\label{Unat}
U(\varphi ) = {U_0}\left[ {1 + \cos \left( {\frac{\varphi }{f}} \right)} \right],
\end{equation}
where $f>0$ is the scale which determines the curvature of the potential. This potential has motivations from string theory and it often arises if the inflaton field is taken to be a pseudo-Nambu-Goldstone boson, i.e. an axion, \cite{Freese1990}. This potential behaves like a small-field potential for $2\pi f < {M_P}$, and like a large-field potential for $2\pi f > {M_P}$ \cite{Baumann2009}. The result of this potential in the standard inflationary scenario is in agreement with the Planck 2015 observational data at 95\% CL, as demonstrated in \cite{Planck2015}.

Although the result of the natural potential (\ref{Unat}) in the standard inflationary scenario is same for the both ranges $0 < \varphi /f < \pi $ and $\pi < \varphi /f < 2\pi $ \cite{Freese1990}, their results are different in the Brans-Dicke theory. To account for this fact, we note that using Eqs. (\ref{phivarphiBD}) and (\ref{VUBD}), the potential (\ref{Unat}) changes under the conformal transformation to the Einstein frame as
\begin{equation}\label{Vnat}
V(\phi ) = {U_0}{e^{ - 2\sqrt {\frac{2}{{2\omega  + 3}}}~\phi }}\left[ {1 + \cos \left( {\frac{1}{f}{e^{\sqrt {\frac{2}{{2\omega  + 3}}}~\phi }}} \right)} \right].
\end{equation}
Consequently, although the natural potential (\ref{Unat}) is symmetric in the Jordan frame, the transformed potential (\ref{Vnat}) is not symmetric in the Einstein frame. Indeed, in the Einstein frame the shape of the potential is not the same for the both sides around the minimum of potential, i.e. for the ranges $\phi  < \sqrt {2\omega  + 3} \ln \left( {\pi f} \right)$ and $\phi  > \sqrt {2\omega  + 3} \ln \left( {\pi f} \right)$. Additionally, these two ranges correspond respectively to the ranges $0 < \varphi /f < \pi $ and $\pi < \varphi /f < 2\pi $, in the Jordan frame. Putting all of these notes together, we conclude that the results of natural potential (\ref{Unat}) are not the same for the two sides of its minimum in the Jordan frame.

\begin{figure}[t]
\begin{center}
\scalebox{1}[1]{\includegraphics{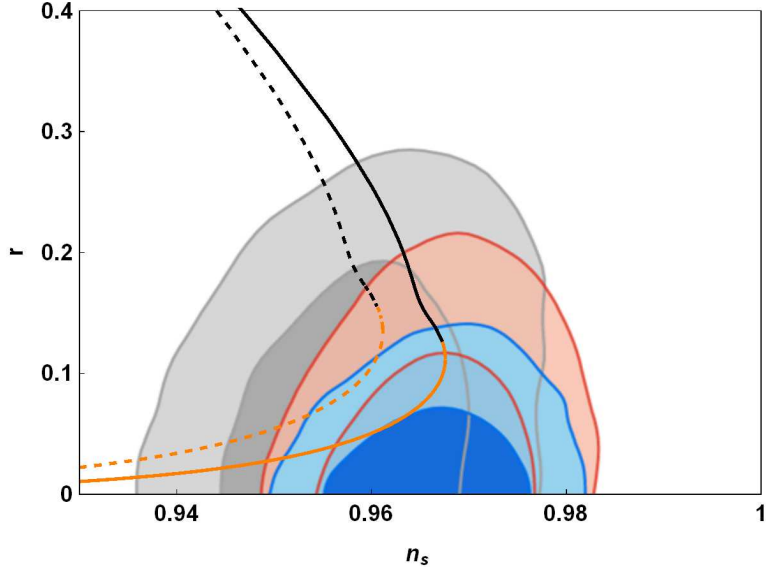}}
\caption{Same as Fig. \ref{fig:pl} but for the natural potential (\ref{Unat}). The predictions of this potential for $N_*=50$ and $N_*=60$ are shown by the dashed and solid black lines, respectively. Furthermore, the results for the two ranges $0 < \varphi /f < \pi $ and $\pi < \varphi /f < 2\pi$ are specified by black and orange colors, respectively.}
\label{fig:nat}
\end{center}
\end{figure}

The results of this potential for the two ranges $0 < \varphi /f < \pi $ and $\pi < \varphi /f < 2\pi $ are shown in Fig. \ref{fig:nat}, by black and orange colors, respectively. We see in the figure that the result of the range  $0 < \varphi /f < \pi $ is outside the region allowed by Planck 2015 TT,TE,EE+lowP data for $N_*=50$, but if we take $N_*=60$, then its result can enter the 95\% CL region of the same data. It is evident from the figure that the result of the range $\pi < \varphi /f < 2\pi$ for both $N_*=50$ and $N_*=60$ can be lied inside the marginalized joint 95\% CL region of Planck 2015 TT,TE,EE+lowP data. It is worth mentioning that in this model, for the case $0 < \varphi /f < \pi $, a larger value of $\omega$ gives a lower value for $r$, while for the case $\pi < \varphi /f < 2\pi$, the greater $\omega$ leads to the smaller $r$. But for the both cases, $r$ approaches $0.16$ ($0.13$) for large values of $\omega$, if we take $N_*=50$ ($N_*=60$). Also, in the both cases, $n_s$ approaches $0.9605$ ($0.9670$), for $N_*=50$ ($N_*=60$).

\subsection{Spontaneously broken supersymmetry (SB SUSY) potential}\label{subsec:SBS}

Here, we proceed to investigate inflation with the spontaneously broken supersymmetry (SB SUSY) potential
\begin{equation}\label{USBS}
U(\varphi ) = {U_0}\left( {1 + b\ln \varphi } \right),
\end{equation}
where $b>0$ is a dimensionless parameter. This potential has wide usage in the hybrid models to provide $n_s<1$ \cite{Mazumdar2011}. However, the result of this potential in the standard inflationary model cannot be compatible with Planck 2015 TT,TE,EE+lowP data \cite{Planck2015}.

\begin{figure}[t]
\begin{center}
\scalebox{1}[1]{\includegraphics{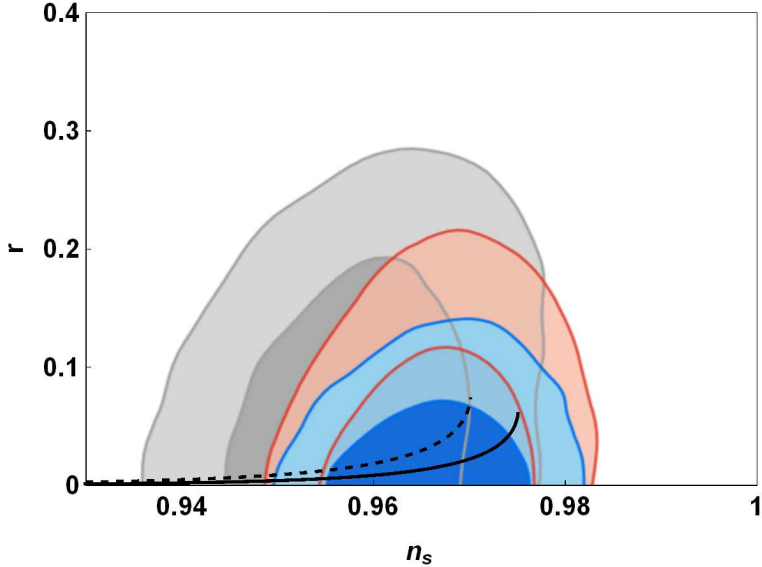}}
\caption{Same as Fig. \ref{fig:pl} but for the SB SUSY potential (\ref{USBS}). The results for $N_*=50$ and $N_*=60$ are demonstrated by the dashed and solid black lines, respectively.}
\label{fig:SBS}
\end{center}
\end{figure}

In the inflationary framework based on the Brans-Dicke gravity, the result of the SUSY breaking potential (\ref{USBS}) for the observational quantities $n_s$ and $r$ depend on the Brans-Dicke parameter and the horizon exit $e$-fold number $N_*$. The result of the potential in $r-n_s$ plane is presented in Fig. \ref{fig:SBS}. It shows that the prediction of the SUSY breaking potential (\ref{USBS}) within the framework of Brans-Dicke gravity in contrary to the standard setting, can lie inside the 68\% CL region of Planck 2015 TT,TE,EE+lowP data \cite{Planck2015}. For this potential, a larger value of the Brans-Dicke parameter $\omega$ gives rise to larger value for both $n_s$ and $r$. But finally, as $\omega$ increases, $n_s$ and $r$ converge respectively to the values $0.9701$ ($0.9751$) and $0.08$ ($0.07$), if we consider $N_*=50$ ($N_*=60$).

\subsection{Displaced quadratic potential}\label{subsec:dq}

The last inflationary potential that we investigate in the Brans-Dicke scenario, is the quadratic potential with displaced minimum
\begin{equation}\label{Udq}
U(\varphi ) = {U_0}{\left( {1 - \frac{\varphi }{\mu }} \right)^2},
\end{equation}
where $U_0$ and $\mu>0$ are constant parameters. By use of the transformation relations (\ref{phivarphiBD}) and (\ref{VUBD}), one can show that for the vanishing Brans-Dicke parameter ($\omega=0$), the above potential changes into the potential corresponding to the Starobinsky $R^2$ inflation in the Einstein frame. Therefore, we can consider the inflationary model with the above potential in the Brans-Dicke gravity as a generalized version of the Starobinsky $R^2$ inflation.

\begin{figure}[t]
\begin{center}
\scalebox{1}[1]{\includegraphics{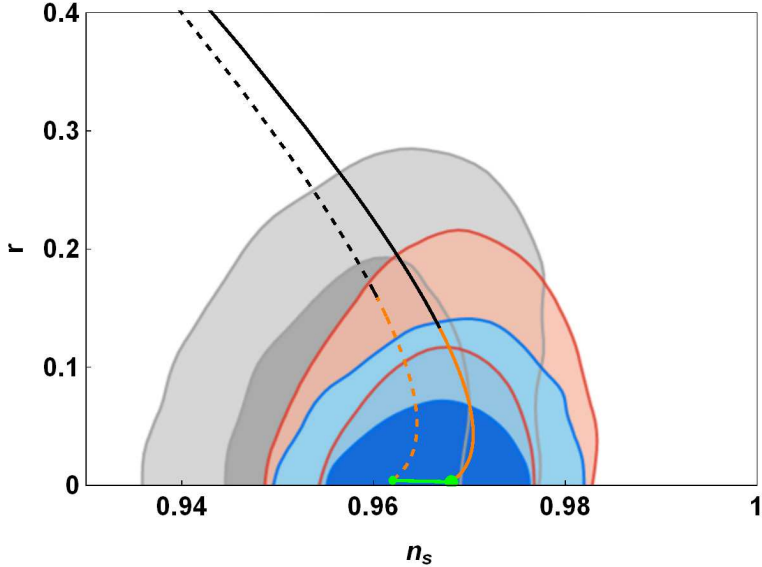}}
\caption{Same as Fig. \ref{fig:pl} but for the displaced quadratic potential (\ref{Udq}). The results for $N_*=50$ and $N_*=60$ are shown by the dashed and solid lines, respectively. Furthermore, the predictions of the model for the two ranges $\varphi  < \mu $ and $\varphi  > \mu $ are specified by black and orange colors, respectively. The result of the Starobinsky $R^2$ inflation for $50 < {N_*} < 60$ is shown by the green line, while the smaller and larger green points demonstrate the results corresponding to ${N_*} = 50$ and ${N_*} = 60$, respectively.}
\label{fig:dq}
\end{center}
\end{figure}

Note that the consistency of the potential (\ref{Udq}) with the Planck 2013 data in the Brans-Dicke gravity has been already investigated by \cite{Tsujikawa2013} using the Einstein frame. But in the present work, we study this potential in the Jordan frame which is our physical frame. Furthermore, we check comparability of this potential in comparison with the Planck 2015 observational data. We show the $r-n_s$ plot of the displaced quadratic potential (\ref{Udq}) in Fig. \ref{fig:dq} in comparison with the observational results. The results of the potential for the ranges $ \varphi  < \mu $ and $\varphi  > \mu $ are specified by black and orange colors, respectively. As it is obvious from the figure, result of the range $ \varphi  < \mu $ is not consistent with Planck 2015 TT,TE,EE+lowP data \cite{Planck2015} for $N_*=50$. But for $N_*=60$, its result can be placed inside the 95\% CL region of the same data. Also, it is clear from the figure that for the range $\varphi  > \mu $, the potential can be in well agreement with the observation such that its prediction can lie inside the 68\% CL region of Planck 2015 TT,TE,EE+lowP data \cite{Planck2015} for both $N_*=50$ and $N_*=60$. We see in Fig. \ref{fig:dq} that the result of the range $\varphi  > \mu $ approaches to the Starobinsky $R^2$ inflation that its prediction has been specified by a green line. In the figure, for the case $\varphi  < \mu $, the larger value of $\omega$ gives rise to a larger value for $n_s$, but a smaller value for $r$. It should be noted that for the both cases $\varphi  < \mu $ and $\varphi  > \mu $, if we take $\omega$ very larger than unity, then the prediction of the potential (\ref{Udq}) for $n_s$ and $r$ converges respectively to the values $0.9604$ ($0.9669$) and $0.16$ ($0.13$), for $N_*=50$ ($N_*=60$).

So far, we tested the predictions of various potentials in $r-n_s$ plane relative to the Planck 2015 observational results. In Table \ref{tab:tab1}, we summarize the results of the examined inflationary potentials. To specify the viable inflationary potentials in light of the observational results, it is further needed to check consistency of their predictions for other inflationary observables such as the running of the scalar spectral index $d{n_s}/d\ln k$, Eq. (\ref{dnsBD}), and the equilateral non-Gaussianity parameter $f_{{\rm{NL}}}^{{\rm{equil}}}$, Eq. (\ref{fNLequil}). We evaluate these two observable parameters for the potentials which are successful in the $r-n_s$ test. Subsequently, we compare our results for different potentials with the results deduced from the Planck 2015 data implying $d{n_s}/d\ln k =  - {\rm{0}}{\rm{.0085}} \pm {\rm{0}}{\rm{.0076}}$ (68\% CL, Planck 2015 TT,TE,EE+lowP) and $f_{{\rm{NL}}}^{{\rm{equil}}} =  - 16 \pm 70$ (68\% CL, Planck 2015 T-only) \cite{Planck2015}. In Table \ref{tab:tab2}, we summarize the predictions of only viable potentials for $d{n_s}/d\ln k$ and $f_{{\rm{NL}}}^{{\rm{equil}}}$ with the allowed ranges for the Brans-Dicke parameter $\omega$, which are compatible with the Planck 2015 results. Here, we should notice that the values of Brans-Dicke parameter $\omega$ in several models do not seem to be compatible with the Solar System constraint, $\omega \gtrsim 10^5$. Of course, if the Brans-Dicke scalar field decays after inflation, the subsequent cosmology coincides with Einstein's general relativity and today’s bound does not apply.

\begin{table*}
	\caption{Results of different inflationary potentials in the Brans-Dicke gravity in $r-n_s$ plane in comparison with the Planck 2015 observational results. Here, the horizon exit $e$-fold number is $N_*=60$.}
	\label{tab:tab1}
	\begin{center}
		\scalebox{1}{
		\begin{tabular}{ |c|c|c|c|c|}
		\hline
		Potential & Standard model & Brans-Dick gravity\\
		\hline
		Power-law & 95\% CL & Outside the region\\
        \hline
		Inverse power-law & Outside the region & Outside the region\\
        \hline
		Exponential & Outside the region & Outside the region\\
		\hline
        Hilltop, $ p=4 $ & 68\% CL & 95\% CL\\
		\hline
		D-brane,  $p=4 $ & 68\% CL & 68\% CL\\
		\hline
        Higgs, $\varphi<\mu $ & 68\% CL & 95\% CL\\
		\hline
		Coleman-Weinberg, $ \varphi<\mu $ & 68\% CL & 95\% CL\\
		\hline
		Natural & 95\% CL & 95\% CL\\
		\hline
		SB SUSY & Outside the region & 68\% CL\\
		\hline
		Displaced quadratic, $ \varphi>\mu $ & 95\% CL & 68\% CL\\
		\hline
	    \end{tabular}
        }
	\end{center}
\end{table*}

\begin{table*}
        \caption{Results of different inflationary potentials in the Brans-Dicke gravity for the running of the scalar index $d{n_s}/d\ln k$ and the equilateral non-Gaussianity parameter $f_{{\rm{NL}}}^{{\rm{equil}}}$. The observables have been evaluated at the horizon exit $e$-fold number $N_*=60$.}
        \label{tab:tab2}
	\begin{center}
		\scalebox{0.8}{
		\begin{tabular}{ |c|c|c|c|c|}
			\hline
            Potential &
			Range of $ \omega $ &
            {Consistency}&
			$ f_{{\rm{NL}}}^{{\rm{equil}}} $  &
            $d{n_s}/d\ln k$ \\
			\hline
            Hilltop, $ p=4 $ &
            ${10^3} \lesssim \omega  \lesssim {10^5}$& 95\% CL &  $ - 0.0054 \lesssim f_{{\rm{NL}}}^{{\rm{equil}}} \lesssim  - 0.0052$  & $ - 0.00019 \lesssim d{n_s}/d\ln k \lesssim  - 0.00016$\\
            \hline
            D-brane, $ p=4$ &
            $900 \lesssim \omega  \lesssim {10^4}$& 68\% CL   & $ - 0.0168 \lesssim f_{{\rm{NL}}}^{{\rm{equil}}} \lesssim  - 0.0074$ & $ - 0.0008 \lesssim d{n_s}/d\ln k \lesssim  - 0.0002$\\
			\hline
            Higgs &
            $2 \times {10^3} \lesssim \omega  \lesssim {10^5}$ & 95\% CL & $ - 0.003 \lesssim f_{{\rm{NL}}}^{{\rm{equil}}} \lesssim  - 0.0005$ & $ - 0.0005 \lesssim d{n_s}/d\ln k \lesssim  - 0.0004$\\
			\hline
            Coleman-Weinberg &
            $700 \lesssim \omega  \lesssim 9000$& 95\% CL  & $ - 0.0058 \lesssim f_{{\rm{NL}}}^{{\rm{equil}}} \lesssim  - 0.0004$  & $ - 0.0004 \lesssim d{n_s}/d\ln k \lesssim  - 2.4 \times {10^{ - 6}}$ \\
			\hline
            Natural &
            $400 \lesssim \omega  \lesssim {10^4}$& 95\% CL  & $ - 0.0385 \lesssim f_{{\rm{NL}}}^{{\rm{equil}}} \lesssim  - 0.0015$  & $ - 0.0023 \lesssim d{n_s}/d\ln k \lesssim  - 3.7 \times {10^{ - 6}}$ \\
			\hline
            SB SUSY &
            $270 \lesssim \omega  \lesssim 4000$& 68\% CL  & $ - 0.0058 \lesssim f_{{\rm{NL}}}^{{\rm{equil}}} \lesssim  - 0.0004$   & $ - 8 \times {10^{ - 5}} \lesssim d{n_s}/d\ln k \lesssim  - 4 \times {10^{ - 7}}$ \\
			\hline
            Displaced quadratic &
            $0 \lesssim \omega  \lesssim 3000$ & 68\% CL & $ - 0.0004 \lesssim f_{{\rm{NL}}}^{{\rm{equil}}} \lesssim 0.0032$ & $ - 0.0006 \lesssim d{n_s}/d\ln k \lesssim  - 0.0004$ \\
			\hline
		\end{tabular}
        }
	\end{center}
\end{table*}

\section{Conclusions}\label{sec:con}

We studied inflation in the framework of Brans-Dicke gravity. For this purpose, first we presented a brief review on the scalar-tensor theories of gravity and expressed the equations governing the background cosmology. We also, reviewed briefly the cosmological perturbations in the scalar-tensor gravity and obtained the scalar and tensor power spectra for this general class of models. Applying the scalar and tensor power spectra, we found relations of the inflationary observables for the model that it makes possible for us to connect theory with observation.

In the next step, we considered the Brans-Dicke gravity as a special case of the scalar-tensor gravity and provided a brief review on this theory of gravity. The Brans-Dicke gravity is based on Mach's principle implying that the inertial mass of an object depends on the matter distribution in the universe so that the gravitational constant should have time-dependence and is usually described by a scalar field. Using the results of the scalar-tensor gravity, we obtained the equations governing the background cosmology in the Brans-Dicke gravity. Then, we considered the slow-roll approximation to simplify the background equations. We further obtained relations of the inflationary observables for the Brans-Dicke gravity, in the slow-roll approximation.

Subsequently, we discussed about the conformal transformations from the Jordan frame to the Einstein frame. Although in this paper, we considered the Jordan frame as our physical frame, however our analysis shows explicitly that in the slow-roll approximation the relations of the inflationary observables including the scalar power spectrum ${{\cal P}_s}$, the scalar spectral index $n_s$ and the tensor-to-scalar ratio $r$, are identical in both the Jordan and Einstein frames.

In addition, we checked viability of different inflationary potentials in the framework of Brans-Dicke gravity. We chose the potentials that have wide usage in study of inflation and they have motivations from quantum field theory or string theory. Our study shows that in the Brans-Dicke gravity, results of the power-law, inverse power-law and exponential potentials lie completely outside the region allowed by the Planck 2015 data, and therefore these inflationary potentials are ruled out. The hilltop, Higgs, Coleman-Weinberg and natural potentials can be compatible with Planck 2015 TT,TE,EE+lowP data at 95\% CL. Moreover, the D-brane and SB SUSY potentials can be in well agreement with the observational data since their results can lie inside the 68\% CL region of Planck 2015 TT,TE,EE+lowP data. Another inflationary potential that we examined in the Brans-Dicke gravity, was the quadratic potential with displaced minimum. This potential for the zero Brans-Dicke parameter ($\omega=0$) leads to the Starobinsky $R^2$ inflation. The result of the quadratic potential with displaced minimum can be placed within the 68\% CL region of Planck 2015 results.

We also examined the other inflationary observables including the running of the scalar spectral index $d{n_s}/d\ln k$ and the equilateral non-Gaussianity parameter $f_{{\rm{NL}}}^{{\rm{equil}}}$ for those potentials whose results in $r-n_s$ plane were consistent with the Planck 2015 data. We concluded that results of those potentials for $d{n_s}/d\ln k$ and $f_{{\rm{NL}}}^{{\rm{equil}}}$ are compatible with the Planck 2015 results too.

\subsection*{Acknowledgements}

The authors thank the referee for his/her valuable comments.






\end{document}